\documentclass[9pt]{article}
\usepackage{spconf,amsmath,graphicx}
\usepackage{amssymb}
\usepackage{enumitem}
\usepackage{multirow}

\usepackage[belowskip=-10pt,aboveskip=2pt]{caption}
\ninept
\newlength\myindent
\def\Width{0\kern2\tabcolsep\ldots\kern1\tabcolsep0}
\usepackage{etoolbox}
\newcommand{\zerodisplayskips}{%
  \setlength{\abovedisplayskip}{3pt}
  \setlength{\belowdisplayskip}{3pt}
  \setlength{\abovedisplayshortskip}{3pt}
  \setlength{\belowdisplayshortskip}{3pt}}
\appto{\normalsize}{\zerodisplayskips}
\appto{\small}{\zerodisplayskips}
\appto{\footnotesize}{\zerodisplayskips}
\title{Lattice-based Improvements for Voice Triggering Using Graph Neural Networks}
\name{Pranay Dighe, Saurabh Adya, Nuoyu Li, Srikanth Vishnubhotla, Devang Naik, Adithya Sagar, Ying Ma, Stephen Pulman, Jason Williams}
\address{Apple, One Apple Park Way, Cupertino, California}
\begin{document}
%
\maketitle
\begin{abstract}
Voice-triggered smart assistants often rely on detection of a trigger-phrase before they start listening for the user request. Mitigation of \textit{false triggers} is an important aspect of building a privacy-centric non-intrusive smart assistant. In this paper, we address the task of false trigger mitigation (FTM) using a novel approach based on analyzing automatic speech recognition (ASR) lattices using graph neural networks (GNN). The proposed approach uses the fact that decoding lattice of a falsely triggered audio exhibits uncertainties in terms of many alternative paths and unexpected words on the lattice arcs as compared to the lattice of a correctly triggered audio. A pure trigger-phrase detector model doesn't fully utilize the intent of the user speech whereas by using the complete decoding lattice of user audio, we can effectively mitigate speech not intended for the smart assistant. We deploy two variants of GNNs in this paper based on 1) graph convolution layers and 2) self-attention mechanism respectively. Our experiments demonstrate that GNNs are highly accurate in FTM task by mitigating $\sim$87\% of false triggers at 99\% true positive rate (TPR). Furthermore, the proposed models are fast to train and efficient in parameter requirements.
\end{abstract}
\section{Introduction}
\label{sec:intro}
Smart assistants are ubiquitous in various devices like smart speakers, mobile phones, smart watches, etc. These devices are controlled by voice commands gated by a trigger phrase or a wake-up word. While the trigger-phrase detection algorithms \cite{sigtia2018vt} are precise and reliable, the operating point may sometimes allow some non-trigger speech or background noise to wake up the device. 
The goal of this paper is to improve the quality of the smart assistant by minimizing the number of false alarms while rejecting a minimal number (ideally zero) of intended triggers.

\begin{figure}[t]
\centering
\includegraphics[width=\linewidth]{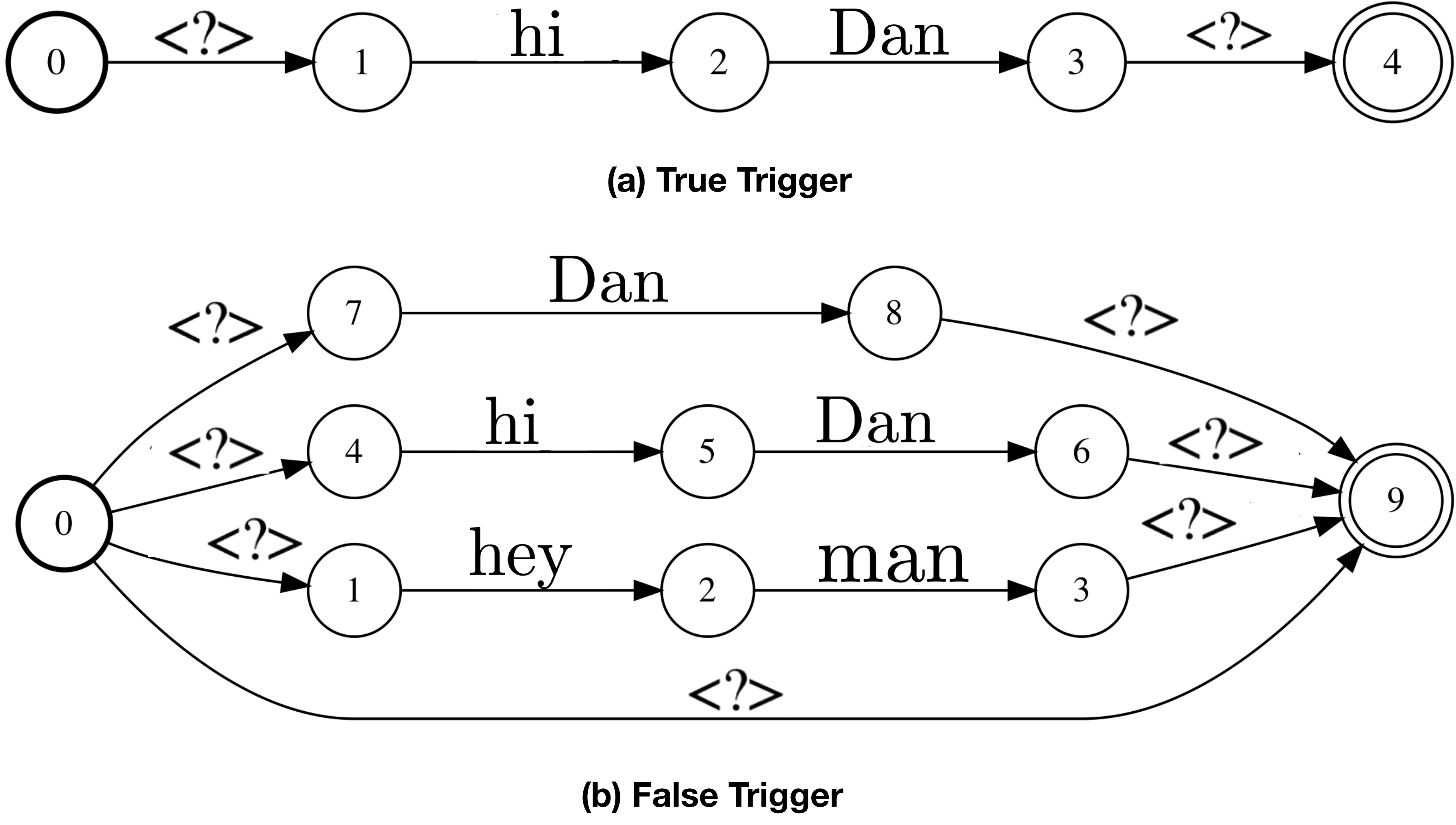}
\caption{Example of true and false trigger lattices.}
\label{fig:lattices}
\end{figure}

The proposed approach utilizes ASR decoding lattices for determining whether a user request is a false trigger or not. Lattices are obtained as Weighted Finite State Transducer (WFST) graphs \cite{mohri2002weighted} during beam-search decoding step in ASR and they concisely represent the top few competing word-sequences hypothesized for the processed utterance. Our FTM approach is based on the hypothesis that a true (intended) utterance spoken by a user is less noisy and the best word-sequence hypothesis has zero (or few) competing hypotheses in the ASR lattice. This is illustrated in a \textit{skinny} lattice shown in Figure \ref{fig:lattices}(a). On the other hand, false triggers often originate either from from background noise or from speech which sounds similar to the trigger-phrase. Multiple ASR hypotheses may compete during decoding in this case and they may be present as alternate paths in the lattices of false trigger utterances. The uncertainty of the ASR decoder could also be exhibited by different distribution of the acoustic model (AM) and language model (LM) scores on the lattice arcs. An example of lattice from a false trigger utterance is shown in Figure \ref{fig:lattices}(b).

\label{sec:features}
\begin{figure*}[t]
\centering
\includegraphics[width=\linewidth]{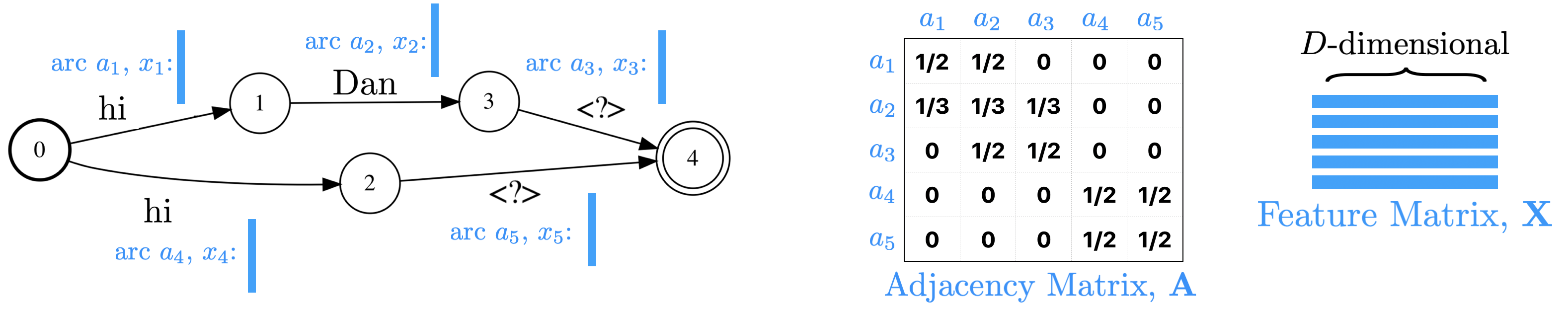}
\caption{Example of lattice represented as adjacency matrix (row-normalized) and feature matrix.}
\label{fig:adj_x}
\end{figure*}

Note that we do not rely on the 1-best ASR hypothesis for FTM here because the acoustic model and language model can sometimes ``hallucinate'' the trigger-phrase. Instead, our approach leverages the whole ASR lattice for FTM. Along with the trigger-phrase audio, we expect to exploit the uncertainty in the post trigger-phrase audio as well. True triggers typically have device-directed speech (e.g.\textit{``hi Dan, what time is it?''} with limited vocabulary and \textit{query}-like grammar whereas falsely triggers may have random noise or background speech (e.g.\textit{``hey Don, let's grab lunch''}). These differences are explicitly exhibited by the decoding lattices and we model them using graph neural networks in this paper. Specifically, we use graph convolution networks (GCN) and self-attention based graph neural networks (SAGNN). The FTM task undertaken in this paper differs from the voice trigger (VT) detection task because we analyze the whole utterance as opposed to a pure trigger detector which focuses on the hypothesized trigger-phrase audio only.

Prior work on FTM mainly comprises of research on key-word spotting and wake-up word detection. VT detection approaches typically rely on multi-stage neural network based processing of acoustic features to determine the presence of the wake-word \cite{sigtia2018vt,kumatani2017direct,wu2018monophone,guo2018time,norouzian2019exploring}. These approaches often use ASR as an auxiliary task to aid the VT detection task. Lattice-based FTM has been successfully explored in \cite{jeon2019latrnn,haung2019study} and has shown that confusion in ASR lattices acts as a strong evidence of false alarm. Prior work on GNNs has been comprehensively summarized in \cite{wu2019comprehensive,kipf2016semi,lee2019self,Zhang2018GaANGA}; and \cite{sperber-etal-2019-self} demonstrates use of self-attention based GNNs on lattice inputs for a machine translation task.

The rest of the paper is organized as follows. Section \ref{sec:features} discusses how to convert lattices into appropriate features for GNN based processing. Section \ref{sec:gnn} explains our GCN and SAGNN based FTM approach. Section \ref{sec:experiments} provides details of FTM experiments, their results and analysis. Finally, Section \ref{sec:conclusions} draws conclusion of this work.

\section{Lattices as Input Features for FTM}
\label{sec:features}
In our experimental system, when a trigger detection mechanism detects a trigger, the system starts processing user audio using a full-blown ASR system. A dedicated algorithm determines the end-of-speech event at which point we obtain the ASR output and the decoding lattice. We use word-aligned lattices such that each arc corresponds to a hypothesized word and derive feature vectors for lattice arcs in a manner similar to \cite{jeon2019latrnn}. 

\begin{figure}[b]
\centering
\includegraphics[width=\linewidth]{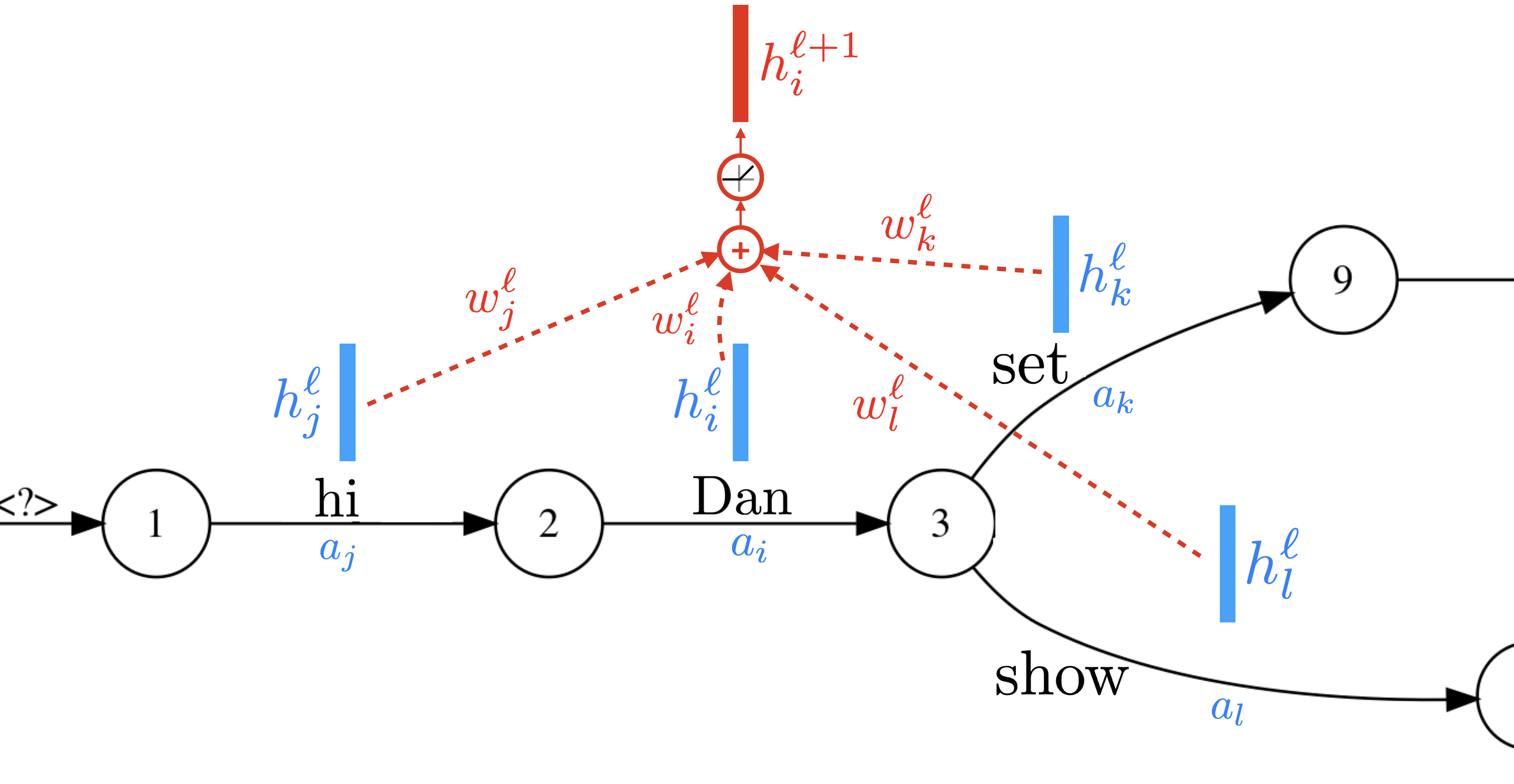}
\caption{Graph convolution operation on a lattice arc aggregates information from all neighbors at a distance on one hop.}
\label{fig:gcn_lattice}
\end{figure}

Lattices can be visualized as directed acyclic graphs defined using a collection of nodes and edges. Denoting lattice arcs as nodes of the graph, there is a directed edge from one node to another if the underlying arcs are connected in the lattice (i.e. the end-state of first arc is the start-state of the second arc). Each node (or arc) has a feature vector associated with it. The FTM task is to take a lattice as a graph input and do a binary classification between true trigger and false trigger class. Formally, if $\mathcal{A}=\{a_1,\hdots,a_{N}\}$ is the set of arcs in the lattice where $N$ is the total number of arcs and each arc $a_i$ has a feature vector $\mathbf{x}_{i} \in \mathbb{R}^{D}$, we can express the lattice in terms of following two matrices:
\begin{itemize}
\item \textbf{Adjacency Matrix}, $\mathbf{A}\in\mathbb{R}^{N\times N}$:  where $A_{ij}=\frac{1}{\text{degree}(a_i)}$ if arc $a_i$ is connected to arc $a_j$ otherwise $A_{ij}=0$, and
\item \textbf{Feature Matrix} $\mathbf{X}\in\mathbb{R}^{N\times D}$: which contains arc features row-wise in the same order as the arcs appear in $\mathbf{A}$. 
\end{itemize}
We design $\mathbf{A}$ to be symmetric so that it captures both forward and backward connections in the lattice. Each arc is also considered as adjacent to itself. An example of a lattice represented as $\mathbf{A}$ and $\mathbf{X}$ is shown in Figure \ref{fig:adj_x}.

\section{Graph Neural Networks for FTM}
\label{sec:gnn}
In this section, we explain the graph convolution and self-attention based GNN architectures for FTM task.

\subsection{Graph Convolution Network}
\label{sec:gcn}

A graph convolution network \cite{kipf2016semi,wu2019comprehensive} differs from a vanilla CNN as it generalizes the 2-D convolution operation to a generic graph structure. For each node in a given graph, the information is accumulated from all its neighbor nodes. The graph does not need to be having a 2-D grid structure (e.g. pixels in an image). Each node in the graph can have different number of neighbors. 

Figure \ref{fig:gcn_lattice} depicts a graph convolution operation using a lattice as an example. The arc $a_i$ has arc $a_j$, $a_k$, and $a_l$ as neighbors. If $h^{\ell}$'s denote the hidden feature vectors associated with the arcs in $\ell^{th}$ hidden layer of a GCN, the graph convolution operation accumulates the information from the neighbors of arc $a_i$ to determine its hidden feature vector $h_i^{\ell+1}$ in the next layer as follows:
\begin{equation}
h_i^{\ell+1} = f(w_i^\ell \mathbf(h_i) + w_j^\ell \mathbf(h_j^\ell) + w_k^\ell \mathbf(h_k^\ell) + w_l^\ell \mathbf(h_l^\ell))
\end{equation} 
where $f$ is a non-linear function and $w^{\ell}$'s denote weight parameters of $\ell^{th}$ layer. Such a graph convolution operation is computed on each arc of the graph simultaneously so that hidden feature vectors of all the arcs are updated. 

 \begin{figure*}[t]
  \begin{center}
  \includegraphics[width=0.7\linewidth]{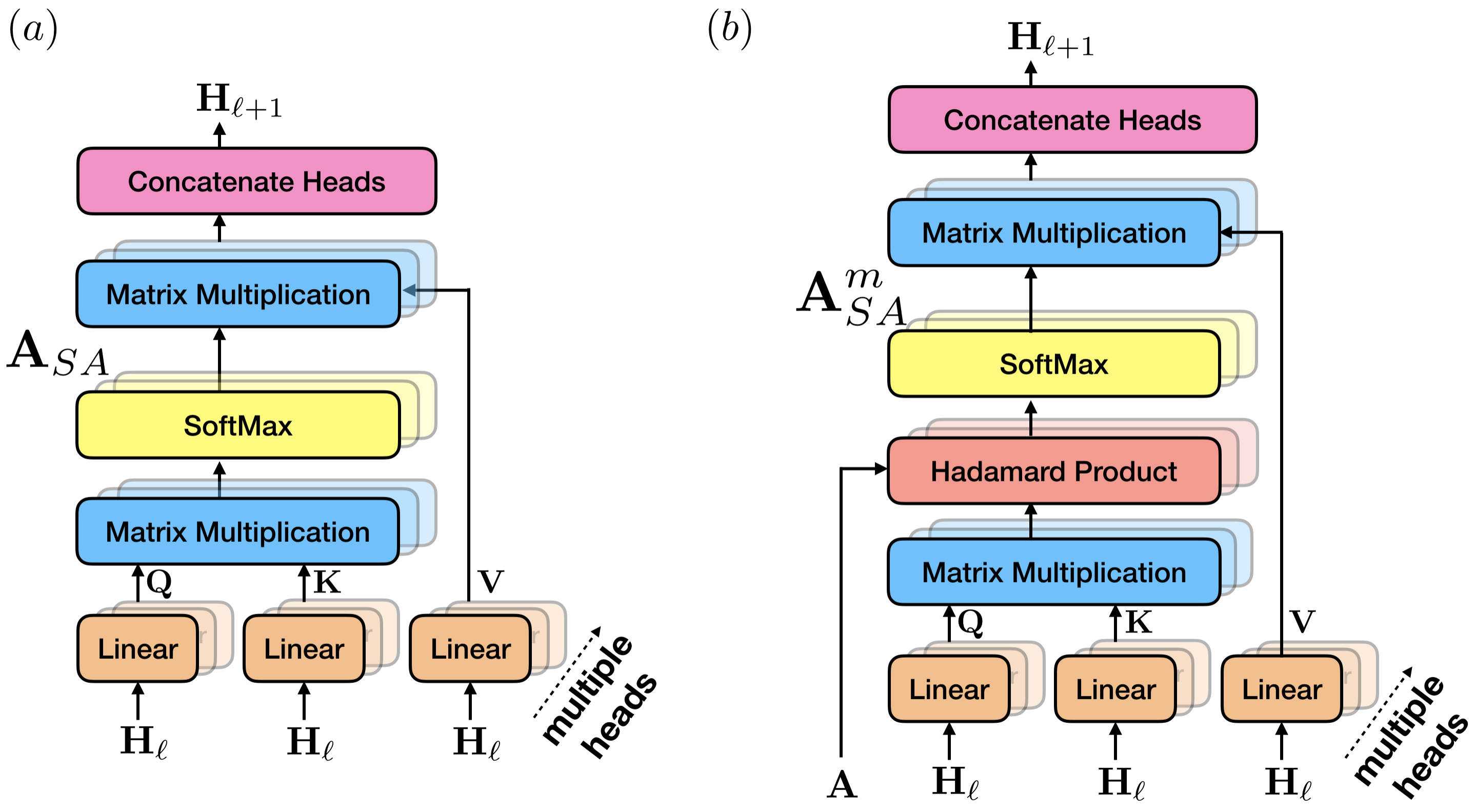}
  \end{center}
\caption{(a) Self-attention GNN layer, (b) Masked Self-attention GNN layer}
\label{fig:gcn_archs}
\end{figure*}

Given a lattice in terms of adjacency matrix $\mathbf{A}$, arc features $\mathbf{X}$,  and a $L$-layer deep GCN with weight matrices $\{\mathbf{W}_{\ell}\}_{\ell=1}^L$ as learnable parameters for each hidden layer with dimensions $\mathbf{W}_{1}\in \mathbb{R}^{D\times H}$ and $\mathbf{W}_{\ell}\in \mathbb{R}^{H\times H}\forall \ell\in [2,L]$, the layerwise transformation is as follows:
\begin{equation}
\mathbf{H}_{\ell+1}=f(\mathbf{A}\mathbf{H}_{\ell}\mathbf{W}_{\ell})
\end{equation}
where $\mathbf{H}_1=\mathbf{X}$, non-linear function $f$ is \textit{ReLU} activation and $\ell\in [1,L]$. The product $\mathbf{H}_{\ell}\mathbf{W}_\ell$ performs a linear transformation of arc features and the product with adjacency matrix imposes the lattice structure using the graph convolution operation. Therefore, $i^{th}$ column of $\mathbf{H}_{\ell+1}$, which corresponding to arc $a_i$'s hidden feature vector, accumulates information from all the neighboring arcs' hidden feature vectors from the previous GC layer. At the output of last GC layer, we have hidden feature vectors for each arc in $\mathbf{H}_{L+1}\in \mathbb{R}^{N\times H}$. We combine the hidden features in $\mathbf{H}_{L+1}$ using average pooling to get an overall lattice-embedding. This embedding is then processed using a fully connected hidden layer followed by \textit{sigmoid} non-linearity to predict the probability of the input lattice being a false trigger. The adjacency matrix $\mathbf{A}$ is fixed for a given lattice and doesn't change for different hidden layers of the GCN. Since $\mathbf{A}$ is row-normalized,  the aggregation of information in GC operation preserves the numerical scale of the hidden feature vectors.

\subsection{Deep-residual Graph Convolution Network}
\begin{figure}[h]
\begin{center}
\includegraphics[width=0.25\linewidth]{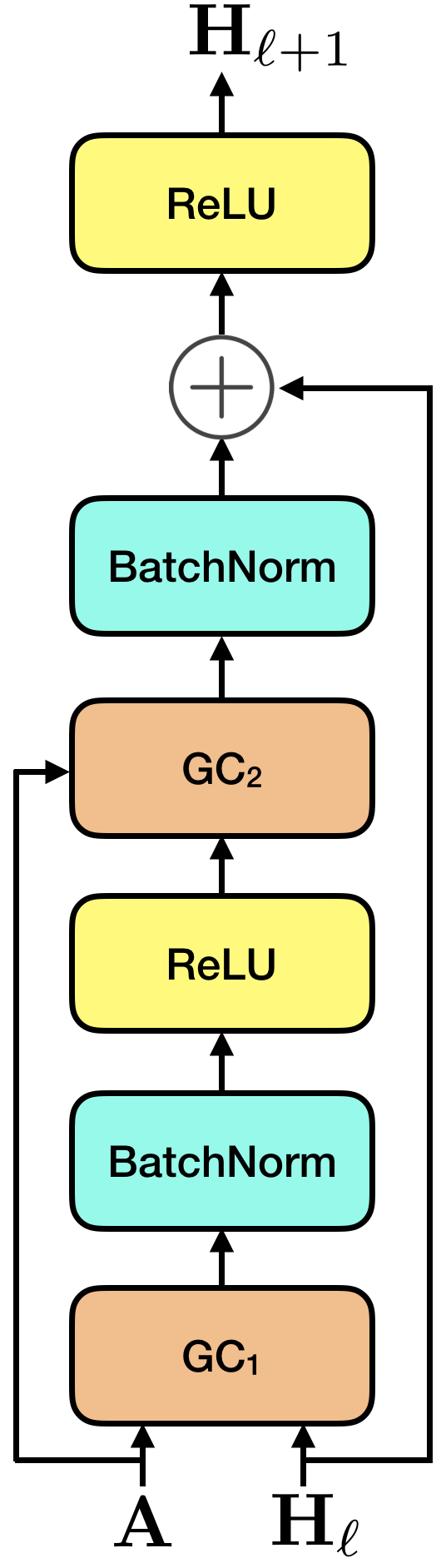}
\end{center}
\caption{Residual block for GCN}
\label{fig:resgcn}
\end{figure}

In each hidden layer in GCN above, the operation $\mathbf{A}\mathbf{H}_{\ell}\mathbf{W}_{\ell}$ aggregates information only from the neighbors at a distance of one hop. With more hidden layers, each arc successively receives information from farther away neighbors in the lattice. Therefore, the depth of the GCN controls how the lattice-level information is accumulated in the lattice-embedding computed at the final GCN layer. In order to train deep GCNs, we use residual connections and batch-normalization layers in our model architecture. Figure \ref{fig:resgcn} shows a residual block containing two GC operations. We stack such residual blocks to train deeper GCNs in this paper.

\subsection{Self-Attention based Graph Neural Network}
\label{sec:sagnn}
Self-attention graph neural networks (SAGNN) \cite{vaswani2017attention,sperber-etal-2019-self,lee2019self,Zhang2018GaANGA} model the relationship between the nodes of a graph using the self-attention mechanism instead of the predefined edges in the graph. Instead of using a fixed adjacency matrix as in GCNs, this approach uses the inner-product of the feature vectors of the lattice arcs to compute their relevance to each other. As the arc features change in successive hidden layers of the network, their mutual relevance with respect to each other changes as well. Therefore, SAGNN can be seen as a generalization of GCNs where the adjacency matrix $\mathbf{A}$ is not static anymore but determined dynamically. 

We use the multi-headed self-attention mechanism \cite{vaswani2017attention} as shown in Figure \ref{fig:gcn_archs}(a). Each head acts as a unique filter to determine \textit{queries} $\mathbf{Q}$, \textit{keys} $\mathbf{K}$, and \textit{values} $\mathbf{V}$ from the arc features. Self-attention computed as $\mathbf{A}_{SA}=\text{Softmax}(\mathbf{Q}\mathbf{K}^{\top})$ is used to linearly combine the \textit{values} together. Outputs of all the heads are then concatenated together to generate the overall output of one SAGNN hidden layer. We use multiple such layers stacked together followed by a final hidden layer for false trigger classification.

Self-attention mechanism ignores the lattice structure completely as $\mathbf{A}_{SA}$ may have non-zero attention values for arc pairs which are not connected to each other or do not appear in the same path in the lattice. To impose the lattice structural constraints, we propose to mask the self-attention matrix $\mathbf{A}_{SA}$ using the lattice's adjacency matrix $\mathbf{A}$ so that the masked self-attention (shown in Figure \ref{fig:gcn_archs}(b)) is defined as $\mathbf{A}_{SA}^{m}=\text{Softmax}(\mathbf{A}\otimes (\mathbf{Q}\mathbf{K}^{\top}))$ where $\otimes$ denotes Hadamard product. Masking ensures that we only aggregate information from arcs which are connected to each other while the aggregation is still weighted using self-attention mechanism. Therefore, masked SAGNN architecture can be seen as a hybrid of GCN and SAGNN approach.

\section{Experimental Analysis}
\label{sec:experiments}

\subsection{Dataset, Features, and Evaluation Metrics}
\label{sec:data}

Table \ref{tab:dataset} summarizes the FTM dataset used in our experiments. For training the models, we augment the \textit{train} and \textit{cv} set to make it 3x bigger by adding gain, noise, and speed perturbations. The input lattice arc feature $x_i$ is a 20-dim vector similar to \cite{jeon2019latrnn} which comprises of 14-dim bag-of-phones embedding from a bottleneck autoencoder, AM score, LM score, arc log-posterior probability, number of frames in the arc, and two binary dimensions which are set as ``1'' if the arc corresponds to the words of the trigger phrase. All models have 64 dimensional hidden state vectors $h_i$ and a binary output which is evaluated using binary cross entropy loss. The models are evaluated on the \textit{eval} set using ROC curves comparing true positive rate (TPR) and false alarm rate (FAR) metrics. We expect our devices to have minimal false alarms and maximum true positives for a good user-experience. Therefore, we focus on the high TPR ($>0.99$) regime in our ROC curves and prefer models which have largest area under the curve (AUC). 

\begin{table}[h]
\small
\centering
\begin{tabular}{lccc}
\hline 
Class & train & cv & eval \\ 
\hline 
True Triggers & 42,675 & 4,742 & 11,646 \\
False Triggers & 18,669 & 2,074 & 11,316 \\
\hline 
\end{tabular} 
\caption{Dataset for false trigger mitigation task.}
\label{tab:dataset}
\end{table}

\subsection{Results and Analysis}
\label{sec:exp_results}

\begin{table}[b]
\small
\centering
\begin{tabular}{lcccc}
\hline 
System & \# of params & AUC & FAR \\ 
\hline 
ASR-Output  & - & - & 0.868 \\ 
BiLRNN & 15,041 & 0.9906 & 0.134\\ 
GCN 6 layers & 26,369 & 0.9888 & 0.185 \\
ResGCN 8 layers & 70,209 & 0.9905 & 0.155 \\
SAGNN 2 layers/4 heads & 39,105 & 0.9906 & 0.150\\
MaskedSAGNN 2 layers/4 heads &  39,105 & 0.9914 & 0.134  \\
\hline
\end{tabular} 
\caption{FTM results using various systems. FAR shown at TPR=0.99 in all systems (except \textit{ASR-Output} where TPR=0.9994).}
\label{tab:results}
\end{table}

We compare the proposed GNN based approaches to two baseline systems: 1) FTM using ASR outputs and 2) the lattice RNN approach from \cite{jeon2019latrnn}. The former is a weak baseline which detects a false alarm if the top ASR hypothesis does not contain the predefined trigger phrase. The latter is a competitive system as it processes the complete lattice using a RNN to detect false alarms.

\begin{figure}[h]
\centering
\includegraphics[width=\linewidth]{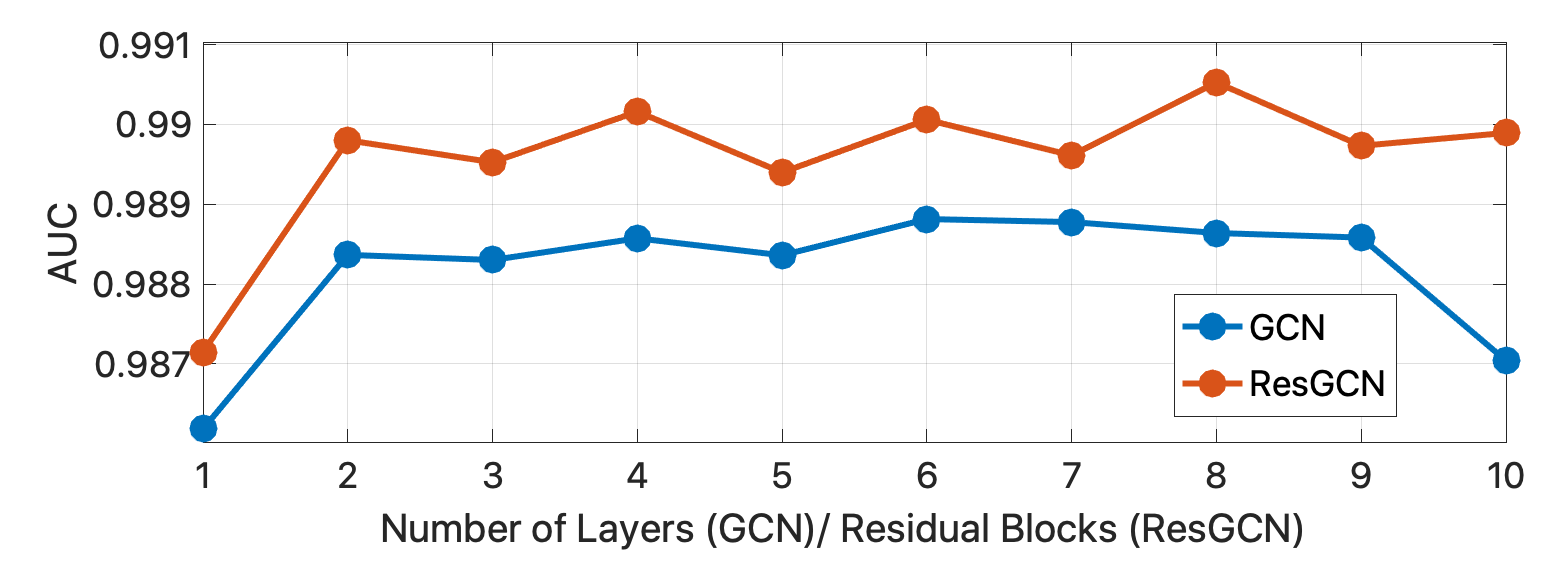}
\caption{AUC for GCN and deep-residual GCN models as a function of number of hidden layers and residual blocks respectively.}
\label{fig:gcn_resgcn_auc}
\end{figure}

Figure \ref{fig:gcn_resgcn_auc} compares various GCN and deep residual GCN models using the AUC metric. We observe that FTM performance of GCN models improves with increasing depth until it starts getting worse after 6 hidden layers. This observation is in line with \cite{li2018deeper,wu2019comprehensive} which explore the difficulty in training deep GCN networks. 

\begin{figure}[h]
\centering
\includegraphics[width=\linewidth]{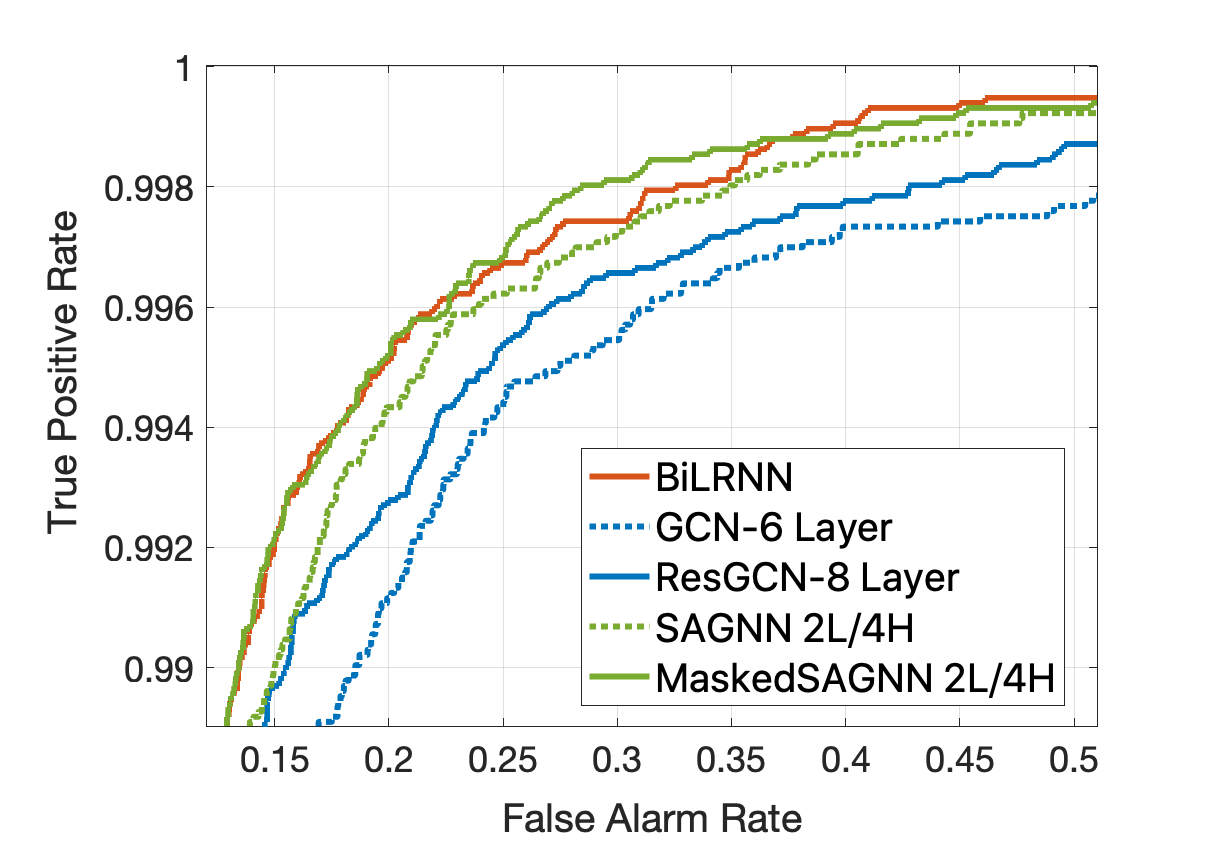}
\caption{ROC curve for GCNs, self-attention GNNs and bi-directional lattice RNN FTM systems.}
\label{fig:sagnn_det}
\end{figure}

We alleviate this issue partially by using residual connections in GCNs (ResGCN) and these models are found to be consistently better than vanilla GCN models. Note that each residual block has 2 GC operations. Therefore, the best ResGCN model with 8 residual blocks has 16 GC operations as compared to 6 GC operations in the best GCN model. ROC curves of best GCN and ResGCN models are shown in Figure \ref{fig:sagnn_det} along with curves for the self-attention GNN models and the baseline bi-directional lattice RNN model. SAGNN and MaskedSAGNN models outperform GCN and ResGCN models conveniently with only 2 hidden layers and 4 attention heads. These models benefit due to the multi-head attention mechanism which can be visualized as \textit{parallelly} acting GC operations. Furthermore, the self-attention values are dynamically learned as opposed to the static adjacency matrix used in GCNs, thereby, focusing \textit{attention} on the relevant parts of the lattice for the FTM task. MaskedSAGNN improves over SAGNN which demonstrates the importance of using lattice structural information in detecting a false trigger. The masking operation may reveal the uncertainty of ASR by exposing the presence of alternative paths in the lattices.

Table \ref{tab:results} further compares our models to the baseline systems. ASR based baseline is not tunable and has a false alarm rate of 0.868. In comparison, lattice RNN baseline and the masked SAGNN approach perform similarly by mitigating $\sim86.6\%$ (FAR=0.134) of false alarms at $1\%$ false rejects (TPR=0.99). Lattice RNNs rely on recurrent operations which result in efficient parameter sharing across the time dimension. While the lattice RNNs perform similar to the GNN-based FTM, we found that they are often inconvenient due to increased training time as compared to the GNNs ($\sim$8min/epoch for RNN training v/s $\sim$1.5min/epoch for GNN training in our experiments). As observed in \cite{sperber-etal-2019-self}, unique connections in lattices inhibit efficient batching of training examples during RNN training. On the other hand, in GNNs, recurrent computations are replaced by graph convolution operations and multiple lattices of different sizes and structures can be efficiently batched together using zero padding resulting in substantial training speed-up.

\vspace{-2mm}
\section{Conclusions}
\vspace{-2mm}
\label{sec:conclusions}
In this paper, we focused on the false trigger mitigation task which improves a voice-triggered smart assistant by making it less intrusive and more privacy preserving. We proposed a novel solution to perform lattice-based FTM using graph neural networks. Lattices are highly informative for distinguishing true triggers from false triggers. We demonstrated that by processing lattices using GNNs, a majority of false alarms can be mitigated at the expense of rejecting few true triggers. We explored two variants of GNNs - namely GCN and self-attention GNN models. We showed that the GCN-based FTM can be improved by increasing network-depth using residual connections. Alternatively, self-attention based GNN models can be improved using adjacency matrix-based masking. We also found that GNN models are fast to train due to efficient processing of lattices using graph convolution operation. In future, we plan to extend GNN-based lattice processing to other tasks such as user-intent classification.

\bibliographystyle{IEEEbib}
\bibliography{refs}

\end{document}